\documentstyle[aps,prc,multicol,epsfig]{revtex}

\begin{document}
\draft
\title{Equation of state and phase transitions in asymmetric nuclear matter}
\author{V.M. Kolomietz$^{1)}$, A.I. Sanzhur$^{1,2)}$, S. Shlomo$^{2)}$ and
        S.A. Firin$^{1)}$}
\address{$^{1)}$Institute for Nuclear Research, Kiev 03680, Ukraine}
\address{$^{2)}$Cyclotron Institute, Texas A\&M University,
College Station, Texas 77843, USA}
\maketitle

\begin{abstract}
  The structure of the 3-dimension pressure-temperature-asymmetry surface
of equilibrium of the asymmetric nuclear matter is studied within the thermal
Thomas-Fermi approximation.
  Special attention is paid to the difference of the asymmetry parameter
between the boiling sheet and that of the condensation sheet of
the surface of equilibrium.
  We derive the condition of existence of the regime of retrograde
condensation at the boiling of the asymmetric nuclear matter.
  We have performed calculations of the caloric curves in the case
of isobaric heating.
  We have shown the presence of the plateau region in caloric curves at the
isobaric heating of the asymmetric nuclear matter.
  The shape of the caloric curve depends on the pressure and is
sensitive to the value of the asymmetry parameter.
  We point out that the experimental value of the plateau temperature
$T\approx 7~{\rm MeV}$ corresponds to the pressure
$P = 10^{-2}~{\rm MeV/fm^3}$ at the isobaric boiling.
\end{abstract}

\bigskip

\begin{multicols}{2}

\section{Introduction}

  At zero temperature, $T=0$, a saturated nuclear matter is stable at
vanishing pressure $P=0$.
  As the temperature increases, the equilibrium state is derived through
the isotherm (equation of state) $P(\rho ,T)$, where $\rho $ is
the particle density.
  In general, there are two different regimes of the equilibrium
state, see Ref. \cite{ll-1}, Ch. 9.
  If the evaporation time is large enough and the evaporated particles are
carried away from the heated liquid, the equilibrium condition is given by
$P(\rho,T)=0$, i.e., it has the same form as at $T=0$.
  By increasing the temperature one can reach the so-called phase separation
temperature $T_{s}$ \cite{arh,bdss,bssd} above which the equilibrium condition
$P(\rho,T)=0$ is not satisfied anymore.
  This happens if the minimum of the isotherm $P(\rho,T)$ is shifted
on the $\rho$-axis and it is located at $\rho=\rho_{{\rm \min }}\neq 0$.
  The temperature $T_{s}$ is then the limiting one for the existence of
a stable state of the liquid phase under the equilibrium condition
$P(\rho,T)=0$.
  In another regime, the heated liquid is surrounded by the saturated vapor.
  The equilibrium condition requires then that the pressure,
$P^{{\rm liq}}(\rho,T)$, and the chemical potential,
$\mu^{{\rm liq}}(\rho,T)$, of the liquid phase should be equal to
the corresponding ones, $P^{{\rm vap}}(\rho,T)$ and
$\mu^{{\rm vap}}(\rho,T)$, for the saturated vapor, see Ref.~\cite{ll-1},
Ch.~8.
  (Here and below indices ``${\rm liq}$'' and ``${\rm vap}$'' denote the
liquid and vapor phases, respectively.)
  In this regime, as the temperature increases, the liquid becomes unstable
with respect to the phase transition (boiling) and the
corresponding boiling temperature is the limiting one for the heating of
the nuclear liquid.

  The limiting temperatures and the critical behavior of the symmetric
nuclear matter have been intensively investigated using an effective 
interaction and finite temperature selfconsistent mean-field theory
\cite{bssd,kwh,scm,jmz}.
  The thermal properties of asymmetric nuclear matter have been considered
in Refs.~\cite{arh,bdss,bssd,lr,lalape,lapera,bb,muse,mekj}.
  In general, the two-component liquid and its saturated vapor co-exist
with different asymmetry parameter $X$ because of the $\rho$-dependence
of the symmetry energy.
  The equilibrium state of the two phases of a two-component system is
described by the 3-dimension surface of equilibrium in variables $P,T$
and $X$, see Ref.~\cite{ll-1}, Chs.~8 and 9.
  The evaluation of the ($P,T,X$)-surfaces of equilibrium and the analysis
of the boiling regime for an asymmetric nuclear matter is a main goal of 
the present paper.
  We also study the properties of the caloric curve in the two regimes 
described above.

\section{Equation of state of hot nuclear matter}

  We will follow the temperature dependent Thomas-Fermi approximation using
the Skyrme-type force as the effective nucleon-nucleon interaction.
  The energy density, ${\cal E}$, and the entropy density, ${\cal S}$,
are given by \cite{br} 
\[
{\cal E}=T\sum_{q}{{\cal A}_{q}^{\ast }J_{3/2}(\eta_{q})}
\]
\[
+{\frac{1}{2}}t_{0}[(1+x_{0}/2)\rho^{2}-
(x_{0}+1/2)(\rho_{n}^{2}+\rho_{p}^{2})] 
\]
\begin{equation}
+{\frac{1}{12}}t_{3}\rho^{\sigma}[(1+x_{3}/2)\rho^{2}-
(x_{3}+1/2)(\rho_{n}^{2}+\rho_{p}^{2})],  \label{e}
\end{equation}
\begin{equation}
{\cal S}=\sum_{q}{({\frac{5}{3}}{\cal A}_{q}^{\ast }J_{3/2}(\eta_{q})-
\eta_{q}\rho_{q}),}  \label{s}
\end{equation}
where $x_{i}$, $t_{i}$ and $\sigma$ are the Skyrme force parameters,
$q$ is the isospin index ($q=n$ for neutron and $q=p$ for proton),
$\rho _{q}$ is the nucleon density and $\rho=\rho_{n}+\rho_{p}$.
  The Fermi integral $J_{\nu}(\eta_{q})=
\int\limits_{0}^{\infty}{dz\,z^{\nu }/(1+\exp (z-\eta_{q}))}$ in
Eqs.~(\ref{e}) and (\ref{s}) depends on the fugacity $\eta_{q}$.
  The value of $\eta _{q}$ can be found from the condition 
\begin{equation}
\rho_{q}={\cal A}_{q}^{\ast}J_{1/2}(\eta_{q}).  \label{cond}
\end{equation}
  Here, ${{\cal A}_{q}^{\ast}=(1/2\pi^{2})}(2m_{q}^{\ast }T/\hbar^{2})^{3/2}$
and $m_{q}^{\ast }$ is the effective nucleon mass derived by 
\[
{\frac{\hbar^{2}}{2\,m_{q}^{\ast}}}={\frac{\hbar^{2}}{2\,m}}
+{\frac{1}{4}}\,[t_{1}\,(1+x_{1}/2)+t_{2}\,(1+x_{2}/2)]\,\rho\]
\begin{equation}
+{\frac{1}{4}}\,[t_{2}\,(x_{2}+1/2)-t_{1}\,(x_{1}+1/2)]\,\rho _{q}.
\label{m}
\end{equation}

  Using Eqs. (\ref{e}) and (\ref{s}) one can derive the free energy density
${\cal F=E-}T{\cal S}$.
  Finally one obtains the pressure $P$ (equation of state) and the chemical
potentials $\mu _{q}$.
  Namely, 
\[
P=\rho^{2}\left( {\frac{\partial}{\partial\rho}}{\frac{{\cal F}}{\rho}}
\right)_{T,X}\ ,
\]
\begin{equation}
\mu_{n}=\left({\frac{\partial{\cal F}}
{\partial\rho_{n}}}\right)_{T,\,\rho_{p}}\ ,\ \ \ \ 
\mu _{p}=\left({\frac{\partial{\cal F}}
{\partial\rho_{p}}}\right)_{T,\,\rho _{n}},
\label{es}
\end{equation}
where the asymmetry parameter $X$ is defined as $X=(\rho_{n}-\rho_{p})/\rho$.
  A numerical calculation of the pressure $P$ with the energy
density ${\cal E}$ from Eq.~(\ref{e}) leads to the van der Waals-like
isotherms $P=P(T,X,\rho)$, describing both the liquid and the vapor phases 
\cite{scm,lr,lalape,lapera}.
  According to Gibbs's phase rule, the equilibrium states of the two-phases
(liquid + vapor) and the two-components (neutrons+protons) system are
located on the 3-dimensional surface in the ($P$, $T$, $X$) space.
  To obtain the shape of the ($P$, $T$, $X$)-surface of equilibrium one
should use the Gibbs equilibrium conditions \cite{ll-1}: 
\[
P^{{\rm liq}}(T,X,\rho)=P^{{\rm vap}}(T,X,\rho )\ ,
\]
\begin{equation}
\mu_{q}^{{\rm liq}}(T,X,\rho )=\mu_{q}^{{\rm vap}}(T,X,\rho )\ .
\label{eql}
\end{equation}
  We will imply that the nuclear matter is a bound system of nucleons
assuming that the following additional condition $\mu_{q}^{{\rm liq}}<0$
is also satisfied.

\section{Surfaces of equilibrium}

  We have performed calculations of the surfaces of equilibrium in
($P$, $T $, $X$) space using the SkM force \cite{br} with the parameters
$t_{0}=-2645$~MeV$\cdot$fm$^{3}$, $t_{1}=385$~MeV$\cdot$fm$^{5}$,
$t_{2}=-120$~MeV$\cdot$fm$^{5}$, $t_{3}=15595$~MeV$\cdot$fm$^{3+3\sigma}$,
$x_{0}=0.09$, $x_{1}=x_{2}=x_{3}=0$ and $\sigma =1/6$, adopted for the
ground state of nuclei.
  We point out that the use of the Skyrme force from the ground-state
calculations is a quite reasonable approximation for our purposes.
  The effective interaction is modified only by a few per cent in a wide
temperature region from 0 to 20~MeV \cite{lgmc,clbl}.
  The numerical procedure we used starts from certain values of
$X^{{\rm liq}}$, $T$ and looks for $\rho^{{\rm liq}}$, $\rho^{{\rm vap}}$
and $X^{{\rm vap}}$ which satisfy the equilibrium conditions (\ref{eql}).

\begin{figure}
\vspace*{0.1in}
\centerline{\epsfxsize=3.3in\epsffile{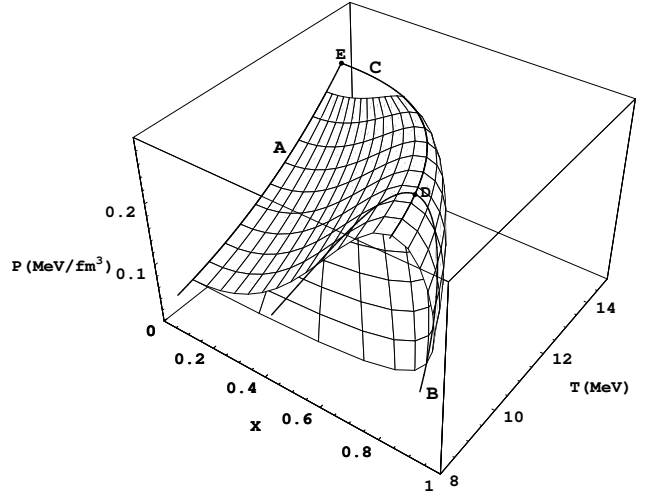}}
\vspace*{0.1in}
\caption{
  Surface of equilibrium in $P,T,X$-space.
  The upper sheet is the surface of boiling and the lower one is the
surface of condensation.
  The azeotropic line and the critical line are marked by the letters A and C,
respectively.
  The drip line $\protect\mu_n=0$ is marked by the letter B.
  The point D indicates the maximum possible asymmetry for the bound liquid
phase at $\protect\mu_n <0$ and the point E is the critical point for
the symmetric system $X=0$.
\label{fig:eqsurf}}
\end{figure}

  The section of the equilibrium surface calculated by our code is shown in
Fig.~\ref{fig:eqsurf}.
  This figure represents the equilibrium states having $X>0$, i.e. states of
the neutron-rich nuclear matter.
  We point out that there are two sheets of the surface of equilibrium.
  The upper sheet is the surface of boiling and the lower one is the surface
of condensation.
  The interior space between sheets is the phase separation region where
the two phases co-exist.
  The crossing points of both sheets with a straight line $P={\rm const}$,
$T={\rm const}$, give the equilibrium asymmetry parameters $X^{{\rm liq}}$
and $X^{{\rm vap}}$ for liquid and vapor phases, respectively.
  In general the asymmetry parameters $X^{{\rm liq}}$ and $X^{{\rm vap}}$
are different. The vapor asymmetry $X^{{\rm vap}}$ exceeds the corresponding
liquid one $X^{{\rm liq}}$ (at $X^{{\rm liq}}>0$).
  This is a feature of the nuclear matter.
  The density dependence of the isospin symmetry energy (see Eq.~(\ref{e}))
provides the condition $|\mu _{n}|<|\mu _{p}|$ and the preferable emission
of neutrons.

  The above mentioned sheets of the surface of equilibrium coincide
along the azeotropic ($P,T$)-line in $X=0$ plane denoted by letter A
in Fig.~\ref{fig:eqsurf}.
  The azeotropic line is cut off at the critical temperature
$T_{{\rm cr}}^{{\rm (sym)}}=14.61$~MeV (point E) of the symmetric
nuclear matter derived by the condition \cite{ll-1} 
\begin{equation}
\left({\frac{\partial P}{\partial\rho}}\right)_{T,X=0}=
\left({\frac{\partial^{2}P}{\partial\rho^{2}}}\right)_{T,X=0}=0\ .
\label{crit1}
\end{equation}
  Both sheets of the surface of equilibrium also coincide along the critical
line marked by letter C in Fig.~\ref{fig:eqsurf}. The critical line of an
asymmetric nuclear matter is derived by the condition, see Ref.~\cite{ll-1},
Ch. 9, 
\begin{equation}
\left({\frac{\partial\mu_{q}}{\partial X}}\right)_{T,P}=
\left({\frac{\partial^{2}\mu_{q}}{\partial X^{2}}}\right)_{T,P}=0\ .
\label{crit2}
\end{equation}
  We point out that the critical temperature $T_{{\rm cr}}$ derived by
Eq.~(\ref{crit2}) at $X\neq 0$\ is different from the one,
$T_{{\rm cr}}^{{\rm (sym)}}$, obtained from Eq.~(\ref{crit1}).
  If one goes along the critical line C, the critical temperature
$T_{{\rm cr}}$ decreases and the corresponding pressure increases as the
value of $X$ increases.

  The line B in Fig.~\ref{fig:eqsurf} is the line with $\mu_{n}=0$.
  The crossing point of the critical line C with the line B (point D
in Fig.~\ref{fig:eqsurf}) provides the maximum possible asymmetry for
the bound ($\mu_{q}^{{\rm liq}}<0$) liquid phase.
  This point is located at $X=0.68$, $T=10.4$~MeV and $P=0.26$~MeV/fm$^{3}$
for the SkM interaction.
  Note that the cold nuclear matter at $T=0$ is bound at $X<0.31$.
  Thus, the hot nuclear matter can exist (in the bound state) at higher
asymmetry than the cold one.
  This feature of hot nuclear matter appears because of the increasing of the
symmetry energy with temperature.

\section{Boiling and caloric curves}

\begin{figure}
\centerline{\epsfxsize=3.3in\epsffile{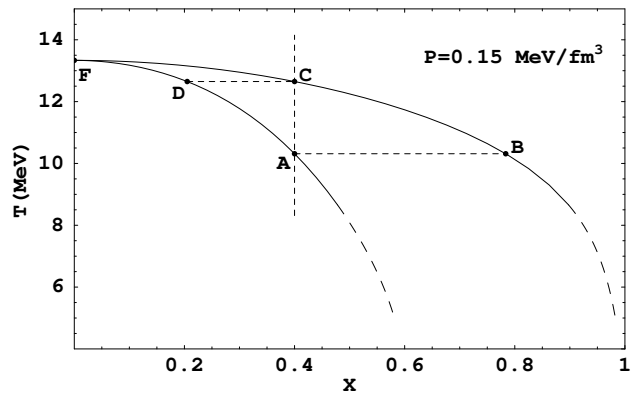}}
\vspace*{0.1in}
\caption{
  A cut of the $(P,T,X)$-surface of equilibrium by the plane
$P=0.15\mbox{~MeV/fm}^3<P_{{\rm cr}}^{{\rm (sym)}}$.
\label{fig:TXdiagr1}}
\end{figure}

\begin{figure}
\vspace*{0.1in}
\centerline{\epsfxsize=3.3in\epsffile{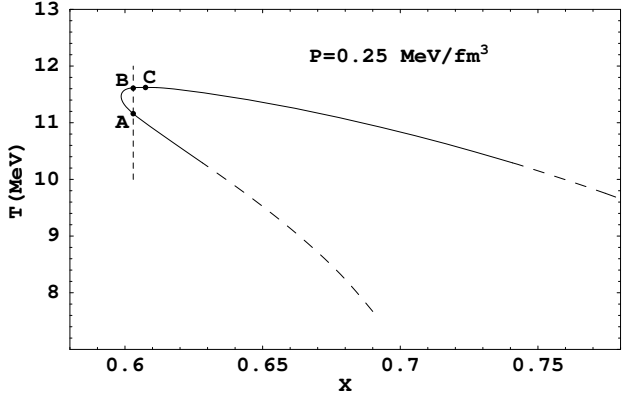}}
\vspace*{0.1in}
\caption{
  The same as in Fig.~{\protect\ref{fig:TXdiagr1}} for
$P=0.25$ MeV/fm$^3$ $>P_{{\rm cr}}^{{\rm (sym)}}$.
\label{fig:TXdiagr2}}
\end{figure}

  Let us consider the isobaric ($T,X$) phase diagrams obtained as
the cut of the equilibrium surface of Fig.~\ref{fig:eqsurf} by plane
$P=P_{{\rm ext}}={\rm const}$.
  These diagrams are very useful for obtaining an insight into
the liquid-vapor phase transition in hot asymmetric nuclear matter.
  The shape of ($T,X$)-diagram depends on the value of the pressure
$P_{{\rm ext}}$.
  The ($T,X$)-diagram contains the critical point $T_{{\rm cr}}$ if the
value of pressure $P_{{\rm ext}}$ exceeds the critical (maximal allowed)
pressure $P_{{\rm cr}}^{{\rm (sym)}}$ on the azeotropic line
in Fig.~\ref{fig:eqsurf}.
  For the SkM force used here, the value of the critical pressure obtained
from (\ref{crit1}) is $P_{{\rm cr}}^{{\rm (sym)}}=$0.2109~MeV/fm$^{3}$.
  In the opposite case $P_{{\rm ext}}<P_{{\rm cr}}^{{\rm (sym)}}$,
the ($T,X$)-diagram contains the point of equal concentration $X=0$.
  In Figs.~\ref{fig:TXdiagr1} and \ref{fig:TXdiagr2} we have plotted the
($T,X$)-diagrams at the pressures $P_{{\rm ext}}=$ 0.15~MeV/fm$^{3}$ and
$P_{{\rm ext}}=$\ 0.25~MeV/fm$^{3}$, respectively.
  In each diagram we have two lines.
  The lower line (on the left) is the line of boiling and the upper line
(on the right) is the line of condensation.
  In Fig.~\ref{fig:TXdiagr1}, these lines meet at the point of equal
concentration ($X=0$) and in Fig.~\ref{fig:TXdiagr2}, they meet at
the critical point $T_{{\rm cr}}$.
  The space between the boiling and condensation lines corresponds to
the states of co-existing phases.

  Let us consider the boiling  of
the asymmetric nuclear matter at the fixed external pressure
$P_{{\rm ext}}<P_{{\rm cr}}^{{\rm (sym)}}$, see Fig.~\ref{fig:TXdiagr1}.
  With the increase of temperature $T$ at a certain asymmetry parameter
$X=X_{{\rm A}}^{{\rm liq}}$, the asymmetric nuclear matter begins to boil at
temperature $T_{{\rm A}}$ on the line of boiling (point A in
Fig.~\ref{fig:TXdiagr1}).
  The boiling process is accompanied by the preferable evaporation of
neutrons because of the condition $|\mu _{n}|<|\mu _{p}|$.
  The liquid phase is then shifted to a more symmetric state and the boiling
temperature increases due to the evaporation of the less-bound particles.
  The final state of system depends on the regime of heating.
  In regime I we assume that the liquid is evaporated in a free space and
the vapor is taken away, the system moves along the surface of equilibrium
to the point F, i.e., reaches the symmetric state
$X=X_{{\rm F}}^{{\rm liq}}=0$ at $T=T_{{\rm \max}}<T_{{\rm cr}}^{{\rm (sym)}}$
on the azeotropic line.
  The boiling is finished then at fixed $T=T_{{\rm max}}$ and $X=0$.
  The asymmetry of the vapor, $X^{{\rm vap}}$, decreases with increasing
$T$ and disappears at $T=T_{{\rm max}}$.
  This means, for instance, that the yield of symmetric clusters at
the condensation of the vapor phase has to grow with $T$.
  In regime II we assume that the vapor remains near the liquid (at fixed 
pressure $P=P_{{\rm ext}}$, this means an evaporation into a closed but not a
fixed volume) the system moves along the boiling line and reaches the point D
in Fig.~\ref{fig:TXdiagr1}.
  The point D corresponds to a fully evaporated liquid at asymmetry
$X_{{\rm C}}^{{\rm vap}}$ of
the vapor (point C on the surface of condensation) that equals to the starting
asymmetry $X_{{\rm A}}^{{\rm liq}}$ of the liquid phase.
  We point out that, in general, the trajectory of motion of the liquid
phase along the surface of equilibrium can be located beyond the plane
$P={\rm const}$ and Fig.~\ref{fig:TXdiagr1} should be considered as
a formal illustration of the behavior of the heating system for the regime
II where both points A and D are taken at different pressures.
  An analogous analysis of the boiling of the asymmetric nuclear matter
can also be done in the case of $P_{{\rm ext}}>P_{{\rm cr}}^{{\rm (sym)}}$,
see Fig.~\ref{fig:TXdiagr2}.
  However the presence of the critical point on the ($T,X$) phase diagrams
in Fig.~\ref{fig:TXdiagr2} leads to a very specific effect of the
retrograde condensation, see also Refs.~\cite{ll-1,muse}.
  If one goes along straight line AB in Fig.~\ref{fig:TXdiagr2} (at
closed volume), the liquid starts to boil at temperature $T_{{\rm A}}$.
  An increase of the temperature leads to an increase of the evaporation.
  However the evaporation begins to decrease at certain temperature
$T<T_{{\rm B}}$ and the vapor disappears at temperature
$T_{{\rm B}}<T_{{\rm C}}$.

  The ($T,X$) phase diagrams allow us to study the shape of the caloric
curve for the case of isobaric heating.
  We will consider the case of the evaporation in a closed (but not fixed)
volume at the fixed pressure $P<P_{{\rm cr}}^{{\rm (sym)}}$, i.e.,
regime II in Fig.~\ref{fig:TXdiagr1}.
  Let us introduce the volume fractions $\lambda^{{\rm liq}}$ and
$\lambda^{{\rm vap}}$ of the liquid and vapor phases defined by
$\lambda^{{\rm liq}}=V^{{\rm liq}}/V$ and $\lambda ^{{\rm vap}}=
V^{{\rm vap}}/V$, where $V^{{\rm liq}}$ and $V^{{\rm vap}}$ are the volumes
of the liquid and vapor phases, respectively, and $V=V^{{\rm liq}}+
V^{{\rm vap}}$.
  The excitation energy per particle, $E^{\ast}/A$, is given by 
\[
\frac{E^{\ast}}{A}={\frac{\lambda^{{\rm liq}}\ {\cal E}^{{\rm liq}}
(\rho ^{{\rm liq}},X^{{\rm liq}},T)
+\lambda^{{\rm vap}}\ {\cal E}^{{\rm vap}}(\rho^{{\rm vap}},
X^{{\rm vap}},T)}{\lambda^{{\rm liq}}\ \rho^{{\rm liq}}+
\lambda^{{\rm vap}}\ \rho^{{\rm vap}}}}
\]
\begin{equation}
-\left( {\frac{{\cal E}^{{\rm liq}}(\rho^{{\rm liq}},
X^{{\rm liq}},T)}{\rho^{{\rm liq}}}}\right)_{T=0}\ ,
\label{ex}
\end{equation}
where ${\cal E}^{{\rm liq}}(\rho^{{\rm liq}},X^{{\rm liq}},T)$ and
${\cal E}^{{\rm vap}}(\rho^{{\rm vap}},X^{{\rm vap}},T)$ are, respectively,
the energy densities of the liquid and vapor phases, derived
by Eq.~(\ref{e}) and taken at the corresponding values of the particle
density $\rho$ and the asymmetry parameter $X$.
  Let us consider a certain asymmetry parameter $X=X_{0}$.
  To calculate the excitation energy $E^{\ast}/A$ we note that
$\lambda^{{\rm liq}}=1$, $\lambda^{{\rm vap}}=0$, and $X^{{\rm liq}}=X_{0}$
if $T$ is below the boiling line and $\lambda^{{\rm liq}}=0$,
$\lambda^{{\rm vap}}=1$ and $X^{{\rm vap}}=X_{0}$ if $T$ is above the
condensation line in the ($T,X$) phase diagram of Fig.~\ref{fig:TXdiagr1}.
  In the case of co-existing phases (space between lines of the boiling
and the condensation in Fig.~\ref{fig:TXdiagr1}) the required values of
$\lambda^{{\rm liq}}$, $\lambda^{{\rm vap}}$, $\rho^{{\rm liq}}$,
$\rho^{{\rm vap}}$, $X^{{\rm liq}}$ and $X^{{\rm vap}}$ can be found using
the following procedure.
  At given $X_{0}$, we go along the straight line $T={\rm const}$ in both
directions and get the cross points with the boiling and condensation lines
which provide the asymmetry parameters $X^{{\rm liq}}$ and $X^{{\rm vap}}$
and the densities $\rho^{{\rm liq}}$ and $\rho^{{\rm vap}}$ for both phases.
  The volume fractions $\lambda^{{\rm liq}}$ and $\lambda^{{\rm vap}}$ are
then determined by the following equations 
\begin{equation}
X_{0}={\frac{\lambda^{{\rm liq}}\rho^{{\rm liq}}X^{{\rm liq}}+
\lambda^{{\rm vap}}\rho^{{\rm vap}}X^{{\rm vap}}}{\lambda^{{\rm liq}}
\rho^{{\rm liq}}+\lambda^{{\rm vap}}\rho^{{\rm vap}}}}\ ,
\ \ \ \lambda ^{{\rm liq}}+\lambda^{{\rm vap}}=1\ .
\label{lambda}
\end{equation}

\begin{figure}
\vspace*{0.1in}
\centerline{\epsfxsize=3.3in\epsffile{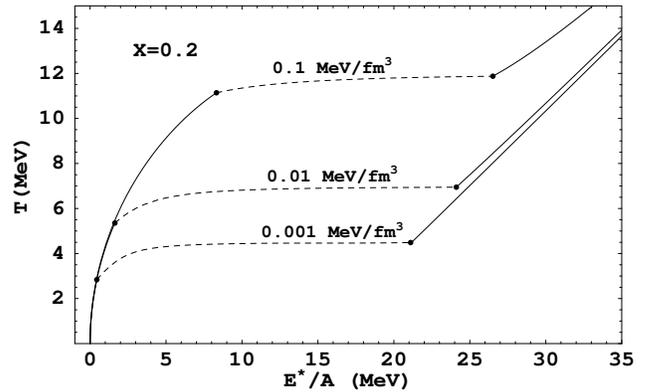}}
\vspace*{0.1in}
\caption{
  Caloric curves for the isobaric heating of the asymmetric nuclear matter
in regime II for different pressure $P$ shown near the curves at fixed
asymmetry parameter $X=0.2$.
\label{fig:calorP}}
\end{figure}

  Caloric curves determined by Eq.~(\ref{ex}) for $X=0.2$ and pressures
$P=10^{-3}$~MeV/fm$^{3}$, $10^{-2}$~MeV/fm$^{3}$ and $10^{-1}$~MeV/fm$^{3}$
are presented in Fig.~\ref{fig:calorP}.
  The solid line at low values of the excitation energy $E^{\ast}/A$
corresponds to the heating of the degenerate Fermi liquid with
$E^{\ast}/A\sim T^{2}$.
  The solid line at high excitation energy $E^{\ast}/A$ describes
the classical Boltzmann's gas with $E^{\ast}/A=(3/2)T$.
  The region of two phase co-existence is displayed by the dashed line.
  The caloric curve is a continuous function of $E^{\ast}/A$ and has a break
in its derivative at two points connecting the two-phase region with
the corresponding single-phase regions.
  As seen from Fig.~\ref{fig:calorP}, the plateau region corresponds
to the two-phase region.
  A small increase of the dashed line with $T$ is due to the motion along
the boiling path AD in Fig.~\ref{fig:TXdiagr1}.
  The value of the plateau temperature increases with the increase of pressure.
  Experimental observations show a nearly flat caloric curve with a
temperature of about 7~MeV \cite{Cibor}.
  If one could assume the process of isobaric heating for the description
of the experimental data, the order of magnitude of the pressure should be
$10^{-2}$~MeV/fm$^{3}$ for this process.
  To obtain more accurate estimation an analysis of the contribution of both
the surface and the Coulomb forces into the equilibrium condition
(\ref{eql}) is needed \cite{mekj,bonch}.

\begin{figure}
\centerline{\epsfxsize=3.3in\epsffile{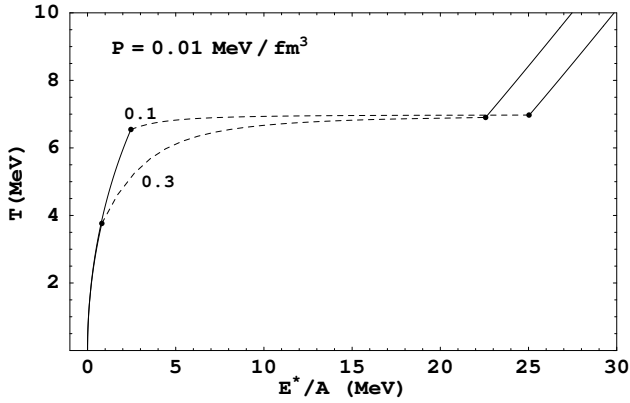}}
\vspace*{0.1in}
\caption{
  The same as in Fig.~{\protect\ref{fig:calorP}} for the fixed pressure
$P=0.01~{\rm MeV/fm^{3}}$ and different values of the asymmetry parameter
$X$ shown near the curves.
\label{fig:calorX}}
\end{figure}

  The asymmetry dependence of the shape of the caloric curve is displayed in
Fig.~\ref{fig:calorX} by plotting two curves at $X=$ 0.1 and 0.3.
  The figure shows that the plateau temperature is slightly sensitive to
the asymmetry parameter.
  At low asymmetry the two-phase region of the caloric curve is flatter
and it is shifted to the lower values of $E^{\ast}/A$ as compared to
the case of high asymmetry.

\begin{figure}
\vspace*{0.1in}
\centerline{\epsfxsize=3.3in\epsffile{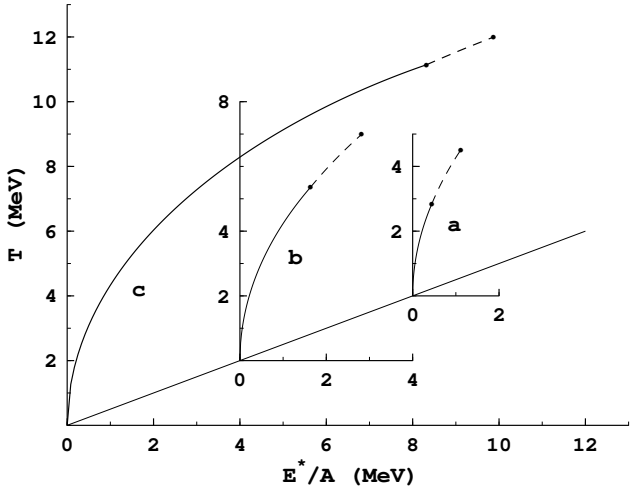}}
\vspace*{0.1in}
\caption{
  Caloric curves for the isobaric heating in the case when vapor is
being taken away from liquid (regime I). Calculations presented for
values of pressure $10^{-3}$~MeV/fm$^3$ (a), $10^{-2}$~MeV/fm$^3$ (b)
and $10^{-1}$~MeV/fm$^3$ (c).
\label{fig:calorO}}
\end{figure}

  Figure~\ref{fig:calorO} shows results of the calculation of the caloric
curve assuming the regime I, i.e. the case when vapor is taken away from
the liquid during the boiling.
  The calculation was carried out for the values of pressure 10$^{-3}$,
10$^{-2}$ and 10$^{-1}$~MeV/fm$^3$.
  The solid lines correspond to the heating of the liquid from $T = 0$ to
$T = T_{{\rm A}}$ with asymmetry parameter $X=0.2$.
  The regions where saturated vapor is present (but taken away) are 
displayed by dashed lines.
  It is seen from Fig.~\ref{fig:calorO} that the value of the external
pressure does not change the shape of the caloric curve calculated for
regime I.
  Only the interval of $E^{\ast}/A$ is sensitive to the value of the
pressure.

\section{Summary}
  Starting from the temperature dependent Thomas-Fermi approximation with the
effective SkM force, we have evaluated the 3-dimensional ($P,T,X$) surface of
equilibrium of the asymmetric nuclear matter.
  The ($P,T,X$) surface of equilibrium contains two sheets which correspond
to the boiling and the condensation.
  Both sheets coincide along the azeotropic line and along the line of
the critical pressure marked by C in Fig.~\ref{fig:eqsurf}.
  In general the vapor asymmetry $X^{{\rm vap}}$ on the surface of
condensation exceeds significantly the corresponding liquid asymmetry
$X^{{\rm liq}}$ (at $X^{{\rm liq}}>0$) lying on the surface of boiling.
  This is due to the fact that the isospin symmetry energy causes
a preferable emission of neutrons.
  The surface of equilibrium is restricted by the drip line (line B
in Fig.~\ref{fig:eqsurf}) where one has $\mu_{n}=0$.
  The ($P,T,X)$ surface of equilibrium in Fig.~\ref{fig:eqsurf} shows also
that the hot nuclear matter exists (in a bound state) at higher asymmetry
than the cold one due to the increase of the symmetry energy with temperature.

  Our analysis of the isobaric ($T,X$) phase diagrams (see
Figs.~\ref{fig:TXdiagr1} and \ref{fig:TXdiagr2}) shows that the process
of isobaric boiling of the asymmetric nuclear matter depends on the
value of the fixed pressure $P_{{\rm ext}}$.
  In the case of low enough pressure
$P_{{\rm ext}}<P_{{\rm cr}}^{{\rm (sym)}}=$0.2109~MeV/fm$^{3}$ the
($T,X$)-diagram contains the point of equal concentration $X=0$ and
the process of boiling is accompanied by a decrease of
the vapor asymmetry with increasing temperature $T$.
  In the case of high pressure $P_{{\rm ext}}>P_{{\rm cr}}^{{\rm (sym)}}$,
the process of boiling leads to the very specific effect of a retrograde
condensation with growing temperature $T$, see Fig.~\ref{fig:TXdiagr2}.

  We have shown in Figs.~\ref{fig:calorP} and \ref{fig:calorX} that the
caloric curve for the process of isobaric heating of the asymmetric
nuclear matter contains a plateau region where both the liquid and
the saturated vapor phases co-exist.
  The position of the plateau region on the caloric curve depends on
the pressure $P_{{\rm ext}}$ and is almost insensitive to a change in
the asymmetry parameter $X$.
  As can be seen from Fig.~\ref{fig:calorX}, the shape of the caloric
curve is significantly changed with asymmetry.
  At low asymmetry the two-phase region of the caloric curve is flatter.
  We pointed out that the experimental observation of the saturation of
the caloric curve at temperature of about 7~MeV is obtained in our
approach as an isobaric boiling of the nuclear liquid at the pressure
$P_{{\rm ext}}\approx 10^{-2}$~MeV/fm$^{3}$.
  Note, however, that the effects of nuclear surface and Coulomb
interaction are not included in our study.
  The asymmetry parameter $X^{{\rm vap}}$ of the vapor phase decreases
along the boiling path of the caloric curve (dashed line in
Figs.~\ref{fig:calorP} and \ref{fig:calorX}).
  Thus, the present model predicts an increase of the yield of
the symmetric clusters at the condensation of the vapor phase with
an increase of the excitation energy $E^{\ast}/A$ in the plateau region
of the caloric curve.

\section*{Acknowledgements}

  This work was supported in part by the US Department of Energy under grant
\# FG03-93ER40773.
  We are grateful for this financial support.
  One of us (V.M.K.) thank the Cyclotron Institute at Texas A\&M University
for the kind hospitality.

\end{multicols}

\end{document}